\documentclass[11pt]{article}
\pdfoutput=1

\usepackage{fullpage}
\usepackage{amsmath}
\usepackage{amssymb}
\usepackage{setspace}
\usepackage{bbm}
\usepackage{dsfont}
\usepackage{graphics}

\usepackage{color}
\usepackage[colorlinks=true]{hyperref} 
\hypersetup{
    bookmarks=true,         
    unicode=false,          
    pdftoolbar=true,        
    pdfmenubar=true,        
    pdffitwindow=false,     
    pdfstartview={FitH},    
    pdftitle={My title},    
    pdfauthor={Author},     
    pdfsubject={Subject},   
    pdfcreator={Creator},   
    pdfproducer={Producer}, 
    pdfkeywords={keyword1} {key2} {key3}, 
    pdfnewwindow=true,      
    colorlinks=true,       
    linkcolor=red,          
    citecolor=blue,        
    filecolor=magenta,      
    urlcolor=cyan           
}
\newcommand{\be}{\begin{equation}}
\newcommand{\ee}{\end{equation}}
\newcommand{\bea}{\begin{eqnarray}}
\newcommand{\eea}{\end{eqnarray}}
\newcommand{\tr}{\textrm{tr}}
\def\bse{\begin{subequations}}
\def\ese{\end{subequations}}

\begin{document}


{
\begin{titlepage}
\begin{center}

\hfill \\
\hfill \\
\vskip 0.75in

{\Large \bf The Area Term of the Entanglement Entropy of a Supersymmetric $O(N)$ Vector Model in Three Dimensions}\\

\vskip 0.2in

{\large  Ling-Yan Hung${}^{a,b,c}$, Yikun Jiang${}^a$, and  Yixu Wang${}^{a}$  }

\vskip 0.3in

{\it ${}^{a}$ Department of Physics and Center for Field Theory and Particle Physics, Fudan University, \\
220 Handan Road, 200433 Shanghai, China} \vskip .5mm
{\it ${}^{b}$State Key Laboratory of Surface Physics and Department of Physics, Fudan University,
220 Handan Road, 200433 Shanghai, China}\vskip .5mm
{\it ${}^{c}$ Collaborative Innovation Center of Advanced  Microstructures,
Nanjing University,\\ Nanjing, 210093, China.}

\end{center}


\begin{center} {\bf ABSTRACT } \end{center}
We studied the leading area term of the entanglement entropy of  $\mathcal{N}=1$ supersymmetric $O(N)$ vector model in $2+1$ dimensions close to the line of second order phase transition in the large $N$ limit. We found that the area term is independent of the varying interaction coupling along the critical line, unlike what is expected in a perturbative theory. Along the way, we studied non-commuting limits $n-1\to 0$ verses UV cutoff $r\to 0$ when evaluating the gap equation and found a match only when appropriate counter term is introduced and whose coupling is chosen  to take its fixed point value. As a bonus, we also studied Fermionic Green's functions in the conical background. We made the observation of a map between the problem and the relativistic hydrogen atom.

\vfill

\noindent \today

\end{titlepage}
}

\newpage

\tableofcontents
\newpage
\section{Introduction}

Entanglement entropy has emerged as a very powerful tool in characterizing important properties of many body systems. It has led to new insights for example in the discovery and classification of new phases of matter, such as, to name a few, these exotic symmetry protected topological phases and topological orders \cite{cglw,Wen:2013oza,KongWen}. Since the beginning, it has been observed that the entanglement entropy of ground states of  local field theories \cite{Sorkin:2014kta, Calabrese:2004eu}, or more generally ground states of local Hamiltonians even in discrete systems, satisfy a so called "area law". The area law is the observation that for some choice of region A in configuration space whose entanglement entropy with its complement one calculates, the leading term in the large region size limit is proportional to the area of the co-dimension one surface bounding region A.  Apart from 1-dimensional systems where there is an exact proof \cite{hastings}, the area law remains a conjecture in other dimensions -- although new insights are emerging more recently that edge toward a complete proof of the statement \cite{Swingle:2014qpa,Swingle:2015ipa}. The emergence of area laws is believed to be profoundly connected to quantum gravity theories, given the similarlities between entanglement entropies and the Bekenstein-Hawking black hole entropy.

There is a series of works that explores how the entanglement entropy, and in particular, the area law changes in the presence of perturbations to some given theories, such as free theories or conformal theories. (It's impossible to exhaust the literature on these topics. See for example \cite{Casini:2005zv,Casini:2005rm, Hertzberg:2010uv, Cardy:2010zs}, which are some of the early papers on the subject).  It is known that the area law term is not universal in the sense that it can have dependence on the precise regularization scheme, and there are some recent effort that extracts universal contributions to the area term from relevant perturbations \cite{Rosenhaus:2014zza,Casini:2014yca}. 

On the other hand, precisely for reasons of generic dependence of regularization schemes in field theories, attention has often been focused on subleading terms in the entanglement entropy, such as the logarithmic terms in even dimensions and the constant terms in odd dimensions, which are known to be scheme independent, and are connected to important characteristics of the underlying theory, such as central charges, or the ``F charge'' in odd dimensions, at conformal fixed points. ( See for example the seminal papers \cite{Casini:2006es, Casini:2012ei, Myers:2010xs,Casini:2011kv} that elaborated these connections. ) These works are generally independent of the details of individual theories, and are based on very general symmetries, such as Lorentz symmetries and conformal invariance. 

It is a curiosity therefore to ask how the entanglement entropy depends on the strength of interaction coupling. For strongly coupled theories, there are very restricted tools at our disposal. We have a plethora of holographic results (For a very recent comprehensive review, see for example \cite{Rangamani:2016dms}). In some supersymmetric theories, a deformed supersymmetry preserving ``entanglement entropy'' can be computed exactly even in strongly interacting theories, first considered in \cite{Nishioka:2013haa}, although attention is not usually paid to the coupling dependence of the area term, if there is a continuous coupling to be tuned in these calculations at all. 

On the other hand, we have large N theories, where entanglement entropies can be computed in the large N saddle point limit for generic interacting couplings. This has been considered near the fixed point in \cite{Metlitski:2009iyg}, and more recently the flow of the entanglement under RG flow of the renormalized mass was obtained, making use of the entanglement first law \cite{Akers:2015bgh}. 

In this paper, we would like to consider another example in which large N techniques would come in useful. We study the $\mathcal{N}=1$ supersymmetric $O(N)$ vector model in 2+1 dimensions. The theory has a critical line that controls phase transitions between the $O(N)$ symmetry preserving phase and the $O(N)$ spontaneously broken phase \cite{Moshe:2003xn}. The virtue of this is that there is a one parameter family of theories sitting on the critical line such that we can study the dependence of the entanglement entropy on this coupling. For simplicity, we will work in the leading large N limit, and compute the entanglement entropy of half-space i.e. $y>0$. We will in particular focus on the area term. As we will see, one issue of interest is that there are various new counter terms that are required as soon as we employ the replica trick and obtain the gap equation there. Not all values of the counter terms can be easily fixed based on physical requirements. The minimal choice would suggest that the Bosonic renormalized mass has no dependence on the coupling, and take exactly the same value as in the Bosonic $O(N)$ vector model \cite{Metlitski:2009iyg}. The Fermionic renormalized mass depends on the inverse of the coupling for any non-zero coupling, and thus do not admit a smooth limit back to the free theory. Surprisingly however, the final form of the area term has no dependence on the coupling. That the area term is rigid in the leading large N limit comes as a surprise, and is possibly an artifact of the large N expansion.

Before we end the introduction however, let us reiterate here why the study of the area term is a well defined question in the current context, even though it is considered in many circumstances as being "non-universal, with cut-off dependence. As discussed in \cite{Liu:2013una}, the entanglement entropy is an expansion in $L/\epsilon$ and $L\mu$ etc, where $L$ is the region size, $\epsilon$ is the UV cutoff and $\mu$ any other mass scales in the theory. The change of the UV cutoff would for example have an interplay with the RG flow. Here, we focus on the entanglement of half-space near the critical line, such that $L\to\infty$. Since we stay on the critical line as $g$ is tuned, there is no further complication of changing the cutoff scheme once it is fixed once at a given $g$.  This should render the physics question we are posing sufficiently well defined.

We will begin with a brief review of the supersymmetric $O(N)$ vector model in section \ref{sec:ON}. Then we will present the details of the computation of the entanglement entropy in section \ref{sec:EE}. 

We will conclude in section \ref{sec:conclude} and relegate some excessive details to the appendix.

\section{Supersymmetric O(N) vector model}  \label{sec:ON}

The  action is given by
\be
S(\phi,\psi) = 1/2 \int d^3x [\partial_\mu \phi \partial^\mu \phi - \mu_\phi^2 \phi^2 + \bar\psi(i \gamma^\mu \partial_\mu - \mu_\psi ) \psi - 2 \frac{g \mu}{N} (\phi^2)^2 - \frac{g^2}{N^2} (\phi^2)^3- \frac{g}{N} \phi^2 (\bar\psi .\psi) - 2 \frac{g}{N}(\phi . \bar\psi)(\phi. \psi)],
\ee
where the bosons $\phi_i$ and fermions $\psi_i$ are in the fundamental representation of $O(N)$, and the Lorentz signature is chosen as (1,-1,-1,-1) here. 

After doing the Wick rotation, introducing the auxilliary fields and integrating out the fermions and bosons fields, the effective action can be written as \cite{Moshe:2003xn, Hung:2013dka}

\begin{equation} \label{effS}
S_{eff}=\int d^3x [-\frac{\lambda \rho}{2}+\frac{g^2\rho^3}{2}+g\mu\rho^2]+ \frac 12 Tr \ln(-\square + \mu_\phi^2+\lambda)-\frac 12 Tr \ln(\partial \!\!\!/+\mu_\psi+g_0 \rho)
\end{equation}

Note that in the above action, $\mu_\phi, \mu_\psi$ and $\mu$ are bare parameters of the theory. When they are equal, the theory preserves supersymmetry, which is the case we will focus on. i.e.
\be\label{susym}
\mu_\phi = \mu_\psi = \mu = \mu_0.
\ee

In the leading large N limit,  the gap equation is given by
\begin{equation}\label{gapequation}
m_\psi = \mu_0 + g \rho, \qquad m_\phi^2 = \mu_0^2 + 4 g\mu\rho + 3 g^2 \rho^2 - g \chi,
\end{equation}
where 
\be
\rho = G_\phi(x,x), \qquad \chi = \tr G_{\psi}(x,x),
\ee
and the trace above refers to the trace wrt spinor indices.
In flat space, these means
\be
G_{\phi}(x,x) = \int \frac{d^3p}{(2\pi)^3}  \frac{1}{p^2 + m_\phi^2}, \qquad \tr G_{\psi} (x,x) = \tr \int \frac{d^3p}{(2\pi)^3 } \frac{ p\!\!\!/  + m_\psi}{p^2 + m_\psi^2},
\ee
We note that $m_\psi$ and $m_\phi$ are physical masses, and therefore take finite values.
However, at $d=3$, both propagators are linearly divergent in the UV.
In fact the linear divergence is given by
\be
\textrm{divergence}(G_\phi) = \frac{\Lambda}{(2\pi)^2} = \frac{1}{2m_\psi}\textrm{divergence}(G_\psi),
\ee
where $\Lambda$ is a UV cutoff.
Therefore this means that the bare couplings $\mu_0$ must in fact be divergent such as to cancel the divergence of $G$ to recover a finite physical mass.

The detailed phase structure in the leading large N limit can be found in \cite{Moshe:2003xn}. 

At criticality, we arrange that
\be
m_\psi =0, \qquad \mu_0 = - g\rho. 
\ee 
Recall that the other gap equation is automatically satisfied after picking the above value for $\mu_0$ for any value of $g$. This means that no extra divergent parameters are needed to remove any further singularities. In fact, it is convenient to compute
\be
m_\phi^2 - m_\psi^2 = 2 g m_\psi \rho - g \chi,
\ee
which clearly shows that the divergence of $\rho$ and $\chi$ cancels each other, leading to finite value of $m_\phi$ as soon as $m_\psi$ is made finite.  Supersymmetry ensures that this in fact vanishes along the supersymmetric preserving saddle points at all masses all the way to $m_\psi = m_\phi=0$.

\section{The area term in the entanglement entropy of half-space} \label{sec:EE}

Having briefly reviewed the theory, we would like to explore its entanglement entropy in this section.
For simplicity, we will consider entanglement of half space i.e. $ y \ge 0$. We will employ the replica trick to extract the entanglement entropy. At replica index $n$, it is equivalent to putting the Euclidean path-integral in a conical space, in which an angle deficit located in the $y-t$ plane is given by $2\pi(n-1)$.  Translation invariance remain intact in the orthogonal direction that we call $x$. The boundary of half space is thus the real line $x$, which has infinite length. We will regulate it only at the end when we extract the area term.

\subsection{Green's function in conical space}
We will collect all the ingredients necessary to recover the entanglement entropy. First, we need to recover the Green's function of both the bosons and fermions in the $n$-replicated space.

As it was observed already in the scalar $O(N)$ model, it is expected that the masses $m_\psi$ and $m_\phi$ would generically acquire $r$ dependence. If we were working with a critical theory at $n=1$, it is then expected purely from dimensional grounds that
\be
m_\phi^2 = \frac{a_n}{r^2}, \qquad m_\psi = \frac{b_n}{r},
\ee
where $a_n, b_n \to 0$ as $n\to 1$.  Currently, we adopt the strategy of computing the gap equation by obtaining $\rho_n, \chi_n$ perturbatively in $n-1$.

\subsubsection{Bosonic Green's function}

In three dimension, the Bosonic Green's function whose mass is dependent on the conical place satisfies
\begin{equation}
\Big{(}\frac{\partial^2}{\partial r^2}+\frac{1}{r}\frac{\partial}{\partial r}+\frac{1}{r^2}\frac{\partial^2}{\partial \theta^2}+\frac{a_n}{r^2}+\frac{\partial^2}{\partial  x^2}\Big{)} G_B(m=\frac{a_n}{r^2},{\boldsymbol{ r}; \boldsymbol{r_1}})=-\delta({\boldsymbol{ r}- \boldsymbol{r_1}})
\end{equation}

The Green's function could be solved by mode expansion, which gives
\begin{equation}
G_B( {\boldsymbol{ r};\boldsymbol{ r_1}})=\sum_{l=-\infty}^{\infty}\frac{e^{i\nu(\theta-\theta_1)}}{2\pi n}\int^{\infty}_{-\infty}\frac{dk_\perp}{2\pi}e^{ik_\perp(x-x_1)}\int_0^\infty kdk\frac{J_{\nu_l}(k r)J_{\nu_l}(k r_1)}{k^2+k^2_\perp}
\end{equation}
in which $\nu_l=\sqrt{\frac{l^{2}}{n^{2}}+a_n}$ for $|l|>0$ and $\nu_l=\alpha_n$ for $l=0$. We define $a_n\equiv{\alpha_n}^2$ with $\alpha_n$ taking either positive or negative values. For $l \neq 0$ we take the positive branch of the solution, whereas at precisely $l=0$ the gap equation appears to force upon us the negative branch of the solution. This issue has been discussed in \cite{Metlitski:2009iyg}, which is related to the threshold of bound state formation.

Now we would like to calculate the leading $n-1$ correction to the Green's function in the conical space.
There are two contributions. First, because of the altered periodicity in the presence of the cone, the Green's function at vanishing mass carries $n-1$ dependence. To linear order in $n-1$, we have

\begin{equation}\label{BGd}
{G_n(\boldsymbol{r;r})}-{G_1(\boldsymbol{r;r})}=-\frac{(n-1)}{32 r}
\end{equation}
which is a special case of $D=3$ of Eq.(4.65) in \cite{Metlitski:2009iyg}.

Now however, there is an extra mass term depending on $a_n$ that carries $n-1$ dependence. 
To compute the leading correction coming from $a_n$, we can treat $a_n$ as a perturbation of the conical space Laplacian. The correction to the Green's function as a power series expansion in $a_n$ is then obtained using 
\be
(\Box_n + \frac{a_n}{r^2}) (G_n + \delta G_n)(\boldsymbol{r};\boldsymbol{r_0}) = -\delta^3(\boldsymbol{r}-\boldsymbol{r_0}).
\ee

Naively therefore, we have
\begin{equation} \label{deltaGB}
\begin{split}
\delta G_n(\boldsymbol{r},\boldsymbol{r}) = &\lim_{\boldsymbol{r''} \to \boldsymbol{r}} -a_n\int d^3 x'\frac{1}{{r'}^2} {G_n(\boldsymbol{r''},\boldsymbol{r^{\prime}})} G_n(\boldsymbol{r^\prime},\boldsymbol{r})\\
=&\lim_{x'' \to x}  -a_n\int\frac{dr^{\prime}}{r^{\prime}}\sum_{l} \frac{1}{2\pi n}\int\frac{dk_\perp}{2\pi} \frac{e^{ik_{\perp} (x-x^{\prime\prime})}}{(k^2+k^2_\perp)(k^{\prime2}+k^2_\perp)} \int^{\infty}_{0} k dk J_{\nu_l}(k r)J_{\nu_l}(k r^\prime)\\
& \times \int^{\infty}_{0} k^\prime dk^\prime J_{\nu_l}(k^\prime r^{\prime})J_{\nu_l}(k^\prime r)\\
=& \lim_{x'' \to x} -a_n\sum_{l} \frac{1}{2\pi n}\int\frac{dk_\perp}{2\pi} \frac{e^{ik_{\perp} (x-x^{\prime\prime})}}{(k^2+k^2_\perp)(k^{\prime2}+k^2_\perp)}\int^{\infty}_{0} k dk J_{\nu_l}(k r)\int^{k}_{0} k^\prime dk^\prime J_{\nu_l}(k^\prime r)(\frac{k^\prime}{k})^{\nu}\frac{1}{\nu_l}\\
=&\lim_{x'' \to x} -a_n\sum_{l} \frac{1}{4\pi n}\int^{\infty}_{0} dk \int^{k}_{0} \frac{k^{\prime}e^{-k(x-x^{\prime\prime})}-k e^{-k^{\prime}(x-x^{\prime\prime})}}{k^{\prime 2}-k^2}dk^\prime J_{\nu_l}(k r)J_{\nu_l}(k^\prime r)(\frac{k^\prime}{k})^{\nu}\frac{1}{\nu_l}\\
=&-a_n\frac{1}{4 \pi^{3/2} n r}\sum_{l}\frac{1}{\nu_l} \int ^{1}_{0}dt\frac{t^{2 \nu_{l}}}{1+t}  {_2}F_1[\frac{1}{2},\frac{1}{2}+\nu; 1+\nu ,t^2]\frac{\Gamma
(\frac{1}{2}+\nu)}{\Gamma(1+\nu)}
\end{split}
\end{equation}
where we take $x=x^{\prime\prime}$ and perform the integral over $k^{\prime}$ only in the last step, and made a change of variables, defining $k^{\prime}=k t$.

In the above calculation, strictly speaking, we should have taken $G_n$ to be evaluated at $a_n=0$. However, supposedly if the expression is regular in $a_n$, then to leading order in $a_n$ it wouldn't have made a difference had we set $a_n \to 0$ in $G_n$ only in the last step. 
We now investigate this limit $\alpha_n\rightarrow 0$. The important surprise is that the $l=0$ term contains a pole in $1/\alpha_n$ inherited from $1/\nu_l$, and therefore that term alone is of $O(\alpha_n)$.
The leading $a_n$ contribution to the Green's function is therefore not linear in $a_n$, but depending on $\sqrt{a_n}$.
Focusing on the $l=0$ term, we finally get
\begin{equation}
\begin{split}
-&a_n\frac{1}{4 \pi^{3/2} n r}\frac{1}{\alpha_n}\frac{\Gamma(\frac{1}{2})}{\Gamma(1)}\int^{1}_{0}dt\frac{1}{1+t}  {_2}F_1[\frac{1}{2},\frac{1}{2}; 1 ,t^2]\\
=&-a_n\frac{1}{4 \pi^{3/2} n r}\frac{1}{\alpha_n}\frac{\Gamma(\frac{1}{2})}{\Gamma(1)}\frac{2}{\pi}\int^{1}_{0}dt\frac{Elliptic K(t^2)}{1+t}\\
=&-\frac{\alpha_n}{16 n r}
\end{split}
\end{equation}

We note immediately that this result is half of that obtained in \cite{Metlitski:2009iyg}. The reason is that our expansion assumed that this is a power series expansion in $a_n$ which however is in fact a function of $\sqrt{a_n}$.  When we computed the linear order term in $a_n$, it is effectively a first derivative of the function subsequently evaluated near $a_n=0$. Now, noting that for $z= x^2$,
\be
\frac{d}{dz}f(x) = \frac{d}{dx}f(x) \times \frac{1}{2 x} ,
\ee
we reckon the factor of two we obtained can be attributed to treating the expansion as a function of $a_n$ when it is in fact a function of $\alpha_n$. Correcting this subtlety, we arrive at

\be \label{GBpole}
\boxed{\delta G_n ({\boldsymbol{ r},\boldsymbol{r}})  =   -\frac{\alpha_n}{8 n r}} \,\,
\ee
recovering correctly the result in \cite{Metlitski:2009iyg}.

The $|l|>0$ terms can in fact be summed, and they evaluate to $-a_n/(16\pi n r)$.

There is an alternative way to think about the pole in $\alpha$ obtained above. If we focus on the $r'\to 0$ limit of the integral, and compute the $k_\perp$ integral first, one can see that the $\nu_0$ term would contribute to a logarithmic divergence in the $r'$ integral precisely if we first take the limit $a_n\to 0$. Therefore, the $\alpha$ pole observed above can also be alternatively be taken as a logarithmic divergence localized at the conical singularity, $r\to 0$. 
To confirm such expectation, let us extract the logarithmic divergence explicitly. One very convenient way is to recall that in Eq.(\ref{BGd}) implies that if we use the Euclidian Green's function $G_0$ to calculate the correction above, the difference would be of order $O(n-1)^3$, assuming that $\alpha_n\sim O(n-1)$.  Let us note that this assumption is supported by evidence in the solution of the gap equation of the $O(N)$ scalar model in \cite{Metlitski:2009iyg}, and also the expectation that the free energy should remain analytic in $n-1$ in the limit $n\to 1$.  As we will see, this assumption is confirmed when we solve the gap equation in our case. So up to $O(n-1)^2$ the result would be the same if we replace $G_n$ by $G_1$.
The massless Bosonic Green's function in 3d Euclidean space is 
\begin{equation}
G_1({\boldsymbol{ r_0},\boldsymbol{r_1} })=\frac{1}{4\pi r_{(3)}}
\end{equation}
in which $r_{(3)}$ is the three dimentional distance which is given by 
\be
r_{(3)}=\sqrt{r_0^2+r_1^2-2r_0 r_1 \cos(\theta_0-\theta_1)+(x_0-x_1)^2}.
\ee

Therefore to extract the leading divergent terms proportional to $a_n$, we can return to (\ref{deltaGB}), and replace $G_n \to G_1$, which gives
\begin{equation}
\begin{split}
&\frac{\alpha_n^2}{16\pi^2}\int^{\infty}_{0}\frac{dr_1}{r_1}\int^{2 \pi n}_0 d\theta_1\int^{\infty}_{-\infty}dx_1(r_0^2+r_1^2-2r_0r_1\cos(\theta_0-\theta_1)+(x_0-x_1)^2)^{-1}\\
=&\frac{\alpha_n^2}{16\pi}\int^{r_0}_{0}\frac{dr_1}{r_0 r_1}\int^{2 \pi n}_0 d\theta_1(1+a^2-2a\cos(\theta_0-\theta_1))^{-1/2}\\
+&\frac{\alpha_n^2}{16\pi}\int^{\infty}_{r_0}\frac{dr_1}{r_1^2}\int^{2 \pi n}_0 d\theta_1(1+a^2-2a\cos(\theta_0-\theta_1))^{-1/2}\\
\approx&\frac{\alpha_n^2}{4\pi r_0}\Big{[}\int_{\frac{\epsilon}{r_0}}^1\frac{da}{a}\textrm{EllpticK}(a^2)+\int_0^1da \textrm{EllpticK}(a^2)\Big{]}\\
=&\frac{\alpha_n^2}{4 r_0}(\log 2-\frac 12 \log \frac{\epsilon}{r_0})
\end{split}
\end{equation}
where we have defined $a \equiv r_1/r_0<1$.

The first term in the second to last line is divergent as $a\rightarrow 0$. Thus we introduce a short range cut-off $\frac{\epsilon}{r_0}$, which shows a logarithmic divergence located at the conical singularity. The coefficient of the logarithmic divergence is given by $-1/(8r)$, precisely that anticipated in (\ref{GBpole}) in the $n\to 1$ limit. 
We note the similarity of this divergence to that observed in \cite{Metlitski:2009iyg} that requires counter term of the form $\int d^3x \delta^2(r) \phi^2 $ localized at the conical singularity.

\subsubsection*{Localized counter terms and conformal fixed points}
Now let's introduce the counter term  $\frac{c}{2}\int d^3x \delta^2(r) \phi^2 $, so that the Green's function would be modified and takes the form 

\begin{equation}
G^c_B({\boldsymbol{ r};\boldsymbol{ r' }})=\frac{1}{-\square_n +\frac{\alpha^2}{r^2}+c\frac{\delta(r)}{r}}
\end{equation}

The calculation of the correction induced by the counter term is straight forward, and it turns out that to the second order in c, the correction is(detailed calculation is displayed in appendix \ref{D})
\begin{equation}\label{counterterm}
\delta_c G^c_B (\boldsymbol{r},\boldsymbol{r}) = -\frac{c}{16 n r_0}+\frac{c^2}{32 r_0}(-\log k \epsilon+\log 2-\gamma)
\end{equation}

We find that one proper form of $c$ that could subtract the divergence can be chosen as
\begin{equation}
c=c_r+\frac{c_r^2}{2}(-\log k \epsilon+\log 2-\gamma)-4\alpha_n^2(\log 2-\frac 12 \log \epsilon/r_0)+\mathcal{O}(n-1)^3
\end{equation}

in which $c_r$ is the coefficient $c$ after renormalization.

So we can immediately see that the beta function for $c_r$ now takes the form

\begin{equation}
\beta(c_r)=\frac{c_r^2}{2}-2\alpha_n^2
\end{equation}

The fixed points for the theory are
\begin{equation}
c_{r\pm}=\pm 2 \alpha_n
\end{equation}

We note that only $c_{r+}$ is a stable fixed point.

There are now two different expressions for the total correction to the Green's function in the replicated space is:
\be
G^c_B(\boldsymbol{ r};\boldsymbol{ r})-G_1(\boldsymbol{ r};\boldsymbol{ r})=   \bigg\{  
\begin{array}{c}
-\frac{\alpha_n}{8 n r_0}-\frac{n-1}{32 r_0}+\mathcal{O}(n-1)^2,\\
-\frac{c_{r+}}{16 n r_0}-\frac{n-1}{32 r_0}+\mathcal{O}(n-1)^2,
\end{array}
\ee
where the first expression is obtained by keeping finite $a_n$, obtaining a finite expression that carries a pole in $\alpha$, whereas the second expression requires regulating the log-divergence when $a_n \to 0$ limit is first taken,  and then counter terms are introduced whose coupling is taken to be one at the stable fixed point.  Reassuringly, these two answers match!

The importance of the introduction of the counter term is already noticed in \cite{Metlitski:2009iyg} and revisited in \cite{Akers:2015bgh}, where it is shown that the boundary term can be understood as following from the conformal coupling of the scalar to the Ricci scalar. Here, it is of note to see that the effect of this term can in fact be replaced by taking the $n\to1$ limit last.

\subsubsection{Fermionic Green's function}

Like bosons, the Fermionic Green's function could acquire a mass term in the conical space as well, such that

\begin{equation}
S_F=(\partial \!\!\!/ +\frac{b_n}{r})^{-1} = \left (
\begin{array}{cc}
\partial_x + \frac{b_n}{r} & - e^{- i \hat{\theta}/n}(\frac{n}{r} \partial_{\hat\theta} + i \partial_r) \\
 - e^{ i \hat{\theta}/n}(\frac{n}{r} \partial_{\hat\theta} - i \partial_r) & - \partial_x + \frac{b_n}{r}
\end{array}\right)^{-1}
\end{equation}
And we define  $G_F$ as
\begin{equation}
\square G_F(r,r_1)=\partial\!\!\!/^2 G_F(r,r_1)=-\delta({\boldsymbol{ r}-\boldsymbol{r_1} })
\end{equation}
which is related to the free Fermionic Green's function by
\begin{equation}
{\partial \!\!\!/}^{-1}={\partial \!\!\!/} G_F.
\end{equation}
Let us clarify here that the coordinates we are using is such that the metric is given by
\be
ds^2 = dr^2 + r^2 d\theta^2 + dx^2 = dx^2 + dy^2 + dz^2, \qquad y= r \cos\theta; \,\,z= r\sin\theta, \,\, \theta = n \hat\theta
\ee
where $\theta$ has periodicity $2\pi n$. Of course, the above coordinates are not  single valued. However, if we are only interested
in evaluating the Green's function at the same point far away from the conical singularity at $ r=0$, space is essentially flat, and this coordinate can be used in a patch by patch fashion, which it is single valued. Then, $G_F$ is solved in one patch, and then transformed to the $r,\hat\theta$ coordinates,  corresponding to patching all the patches together to recover one single valued Green's function. 

In the $n$-replicated space therefore, the Fermionic Green's function has a mode expansion as follows:
\begin{equation}\label{Gf}
\begin{split}
G_F(\boldsymbol{ r},\boldsymbol{r_1})&=\sum_{l=-\infty}^{\infty}\frac{e^{i\nu_l(\theta-\theta_1)}}{2\pi n}\int^{\infty}_{-\infty}\frac{dk_\perp}{2\pi}e^{ik_\perp(x-x_1)}\int_0^\infty kdk\frac{J_{\nu_l}(k r)J_{\nu_l}(k r_1)}{k^2+k^2_\perp}\\
&=\sum_{l=-\infty}^{\infty}\frac{e^{i\nu_l(\theta-\theta_1)}}{2\pi n}\int^{\infty}_{-\infty}\frac{dk_\perp}{2\pi}e^{ik_\perp(x-x_1)}I_{\nu_l}(k_{\perp} r)K_{\nu_l}(k_{\perp} r_1)\,(  \textrm{for } \, r<r_1) .
\end{split}
\end{equation}
Here, $\nu_l=|\frac{2l+1}{2n}|$, so that the anti-periodic boundary condition is satisfied around the $\theta$ circle. 

Just like the Bosonic green's function, we tried to get the whole spectrum of the fermion (in order to calculate the Renyi entropy). And we found that it is closely related to the hydrogen atom, which is shown in appendix \ref{B}. However, the significant difference from the hydrogen atom is that the potential is an imaginary one, so the current problem does not have bound state solutions, unlike the hydrogen atom. However, the scattering problem for the hydrogen atom does not have a rigorous analytic expression up to our limited knowledge. But if we only calculate entanglement entropy, only the order $(n-1)$ terms are important, which leads to the strategy similar to the case of bosons here, we use the $(n-1)$ expansion to obtain the leading order corrections of the propagator from $b_n$
\begin{equation}
\begin{split}
\delta S_F(\boldsymbol{ r_0},\boldsymbol{r_1}) =&-\frac{1}{\partial \!\!\!/}\frac{b_n}{r}\frac{1}{\partial \!\!\!/}=-\partial \!\!\!/G_F\frac{b_n}{r}\partial \!\!\!/G_F\\
&=-\int r_2dr_2\int dx_2\int d\theta_2 \partial \!\!\!/_0 G(r_0,r_2)\frac{b_n}{r_2}\partial \!\!\!/_2 G(r_2,r_1)\\
&=b_n\int dr_2\int dx_2\int d\theta_2 \partial \!\!\!/_2 G(r_0,r_2)\partial \!\!\!/_2 G(r_2,r_1).
\end{split}
\end{equation}
Taking the trace of the above expression , we get
\begin{equation}\label{fermioncorrection}
\begin{split}
\tr \delta S_F(\boldsymbol{r_0},\boldsymbol{r_1}) = &2b_n\int dr_2\int dx_2\int d\theta_2 \partial_{x_2} G(r_0,r_2)\partial_{x_2} G(r_2,r_1)+\partial_{r_2} G(r_0,r_2)\partial_{r_2} G(r_2,r_1)+  \\
& \frac{1}{r_2^2}\partial_{\theta_2} G(r_0,r_2)\partial_{\theta_2} G(r_2,r_1)\\
=&-2b_n\int dr_2\int dx_2\int d\theta_2 \big{(} G(r_0,r_2)\square G(r_2,r_1)-G(r_0,r_2)\frac{\partial_{r_2}}{r_2} G(r_2,r_1)\big{)}\\
=&2\frac{b_n}{r_1}G(r_0,r_1)+2b_n\int dr_2\int dx_2\int d\theta_2 G(r_0,r_2)\frac{\partial_{r_2}}{r_2} G(r_2,r_1)
\end{split}
\end{equation}
Now we focus on the second term in Eq (\ref{fermioncorrection}), and take $\boldsymbol{r_0}\rightarrow\boldsymbol{r_1}$ hereafter.
\begin{equation}\label{fcorrectione}
\begin{split}
&2b_n\int dr_2\int dx_2\int d\theta_2 G(r_0,r_2)\frac{\partial_{r_2}}{r_2} G(r_2,r_0)\\
=&\frac{b_n}{2\pi^2 n}\sum_{l=-\infty}^{\infty}\int^{\infty}_{-\infty}dk_\perp\Big{(}\int^{r_0}_{0}\frac{dr_2}{r_2}I_{\nu}(k_{\perp} r_2)K^2_{\nu}(k_{\perp} r_0)\partial_{r_2}I_{\nu}(k_{\perp} r_2)+\int^{\infty}_{r_0}\frac{dr_2}{r_2}K_{\nu}(k_{\perp} r_2)I^2_{\nu}(k_{\perp} r_0)\partial_{r_2}K_{\nu}(k_{\perp} r_2)\Big{)}\\
=&\frac{2b_n}{\pi^2 n r_0^2}\sum_{l=1}^{\infty}\frac{1}{2(4\nu^2-1)}+\frac{b_n}{4\pi^2 r_0^2}(1-2\log(\epsilon/r_0))-\frac{b_n}{2\pi^2 r_0^2}\\
=&-\frac{b_n}{2\pi^2 r_0^2}\log(\epsilon/r_0)+ b_n \times \mathcal{O}(n-1) 
\end{split}
\end{equation}

The $\epsilon$ is a cut-off in the distance to the conical singularity. Much like what happens in the case of bosons, the calculation is strongly suggestive of the need to include a counter term that is localized at $r\to 0$. However, the term  of the form 
$\gamma \int d^3 x \delta(r) \bar\psi \psi$ is such that $\gamma$ is dimensionful, and thus would not produce a divergence term that behaves in the same way as the one observed above. The above calculation is suggestive of a non-local counter term perhaps of the form
$\int \bar\psi {\partial \!\!\!/}^{-1} \partial_r {\partial \!\!\!/}^{-1} \psi$. Without knowing the precise form of the counter term, we resort to a different strategy. Much like what happens for bosons, we can compute  the above correction keeping $n$ general, and taking the $n\to1$ limit only at the end. The details of this calculation is relegated to appendix \ref{B}. In that case, the integral is finite, and we obtain
\be
 -\frac{b_n}{2\pi^2 r_0^2 (n-1)} +\frac{b_n g \,(n-1)}{24 r^2} + \mathcal{O}(n-1)^2,
\ee
where we find no $\mathcal{O}(n-1)^0$ term exactly as in (\ref{fcorrectione}) above when a cutoff was introduced.

%

\subsection{The gap equation in replicated space in $n=1$ expansion}

We now turn to the gap equations (\ref{gapequation}). We may expect that the interaction would break the supersymmetry, so we restore the notation to show possible deviation from the supersymmetric critical line:

\begin{equation}
m_\psi = \mu_\psi + g_{\psi} \rho, \qquad m_\phi^2 = \mu_\phi^2 + 4 g_{\phi_1}\mu\rho + 3 g^2_{\phi_2} \rho^2 - g_{\psi} \chi,
\end{equation}
in which $\mu_\psi$ and $\mu^2_\phi$ are the bare masses of the fermion and boson. Like in the flat $n=1$ case, these bare masses need to be renormalized, absorbing divergences in $\rho$ and $\chi$. To make these precise, we consider computing $\rho$ and $\chi$ using Pauli-Villars regularization so that the divergences can be isolated clearly. 

\subsubsection{Pauli-Villars regularization of the Bosonic and Fermionic propagator}
We would like to use Pauli-Villars\footnote{We thank S. Sachdev for sharing his notes explaining how this is done in replicated space.} regularization to modify the Euclidian Green's function a little. The idea is to consider a modified propagator which is the original propagator subtracted by one corresponding to a boson/fermion with a mass $M$. i.e. Replace
\be
G_{F/B} \to G^{R}_{F/B} = G_{F/B} (M=0)- G_{F/B}(M).
\ee  
In the limit $M\to\infty$, this extra term $G_{F/B}(M)$ approaches zero. We will keep the mass $M$ finite, and take $M\to \infty$ only at the end. 

Under mode expansion, the regulator for fermion now takes the form of 
\begin{equation}\label{Gf}
\begin{split}
G_F(M)&=\sum_{l=-\infty}^{\infty}\frac{e^{i\nu(\theta-\theta_1)}}{2\pi n}\int^{\infty}_{-\infty}\frac{dk_\perp}{2\pi}e^{ik_\perp(x-x_1)}\int_0^\infty kdk\frac{J_\nu(\sqrt{k^2+M^2} r)J_\nu(\sqrt{k^2+M^2} r_1)}{k^2+M^2+k^2_\perp}\\
&=\sum_{l=-\infty}^{\infty}\frac{e^{i\nu(\theta-\theta_1)}}{2\pi n}\int^{\infty}_{-\infty}\frac{dk_\perp}{2\pi}e^{ik_\perp(x-x_1)}I_{\nu}(\sqrt{k_{\perp}^2+M^2} r)K_{\nu}(\sqrt{k_{\perp}^2+M^2} r_1)(r<r_1) 
\end{split}
\end{equation}

Using the uniform expansion for both order and argunment of the Bessel functions\cite{Subir}
\begin{equation}
I_{\nu}(k r)K_{\nu}(k r)\approx\frac{1}{2\sqrt{k^2r^2+\nu^2}}.
\end{equation}
The regulated Fermionic Green's function at replica index $n$ then takes the form
\begin{equation}
G^R_{F\,\,n}(r;r)=\frac{M_{\psi}}{4\pi}-\frac{1}{4\pi^2 n r}\Big{(}\log(2\cosh \pi)+\sum_{l}(\psi(|\frac{l+1/2}{n}|+\frac 12)-\frac 12 \log((\frac{l+1/2}{n})^2+1))\Big{)}=\frac{M_{\psi}}{4\pi}-\frac{c_F(n)}{4\pi^2 n r}
\end{equation}
in which $\psi(x)$ is the polygamma function, and $c_F (n)$ is a constant that depends on n and in the first order of $（n-1）$ takes the form of $c_F (n)=1+\frac 12 (n-1)$. It is noteworthy that this term is non-vanishing in the limit $n\rightarrow1$. 

Taking the same strategy, the regulated Bosonic Green's function is given by

\begin{equation}
G^R_{B\,\,n}(r;r)=\frac{M_{\phi}}{4\pi}-\frac{1}{4\pi^2 n r}\Big{(}\log(2\sinh \pi)+\sum_{l}(\psi(|\frac{l}{n}|+\frac 12)-\frac 12 \log((\frac{l}{n})^2+1))\Big{)}=\frac{M_\phi}{4\pi}-\frac{c_B(n)}{4\pi^2 n r}
\end{equation}
in which $c_B(1)$ is exactly 0 and $c_B(n)=(n-1)\pi^2/8$ to the leading order of $(n-1)$. We note that this result is consistent with Eq.(\ref{BGd}).

\subsubsection{Renormalization of the gap equation}

In Euclidian space, as pointed out in Eq.(\ref{susym}) all the mass scales coincide with each other when supersymmetry is exact, as well as all the coupling constants. So we may expect that in the replicated space the masses would deviate from each other whose value is proportional to $(n-1)$ or higher order in $(n-1)^2$. So in the renormalization process we put in the ansatz
\begin{equation}
\mu_{\phi}^2=\mu_{\psi}^2+A, \quad \mu=\mu_{\psi}+B
\end{equation}
while $\mu_{\psi}$ is chosen such that the first gap equation is properly renormalized.

We find that (taking $n\to 1$ last in the fermion integral to avoid the logarithmic divergence)
\begin{eqnarray}
\mu_{\psi}&=&\frac{b_n}{r}-g\rho=-\frac{M}{4\pi}\\
A&=&0\\
B&=&-\frac{b_n}{2r}
\end{eqnarray}
Indeed in the limit $n \rightarrow 1$ , $A$ and $B$ vanish as $b_n$ vanishes, so that flat space supersymmetry is recovered.

So after the renormailzation, the gap equations now take the form of
\begin{eqnarray}\label{gapeqR}
\frac{b_n}{r}&=&g(-\frac{\alpha_n}{8 n r}-\frac{(n-1)}{32 r})\\
\frac{a_n}{r^2}&=&\frac{ b^2_n}{r^2}-\frac{g b_n}{2\pi^2 r^2(n-1)} +\frac{b_n g \,(n-1)}{24 r^2} + b_n \mathcal{O}(n-1)^2
\end{eqnarray}

\subsubsection{Solution of the gap equations}

A set of self consistent solution perturbatively in $n-1$ is 
\begin{eqnarray}\label{sloution1} 
\alpha_n&=&-\frac{(n-1)}{4},\\
b_n&=&-\frac{\pi^2(n-1)^3}{8g_{\psi}}.
\end{eqnarray}

We find that $a_n$ is independent of all the coupling constants, which is exactly the same number as   (6.27) in reference \cite{Metlitski:2009iyg}. This is not surprising since the Fermionic part is just an order $(n-1)^3$ one. On the other hand, to the lowest order, $b_n$ is inversely proportional to $g_{\psi}\equiv g$, which is a manifestation of the non-perturbative nature in the large $N$ calculation. This set of solution do not admit a smooth limit back to $g\to 0$.  At precisely $g=0$, the only solution is $a_n=b_n=0$, as expected of a non-interacting theory.

\subsection{The area term }

The entropy takes the form of
\begin{equation}
S_{EE}= \partial_n S_{\textrm{eff}}(n)\vert_{n=1}-  S_{\textrm{eff}}(1)
\end{equation}
With a little bit rearrangement, Eq.(\ref{effS}) now takes the form of 
\begin{equation}
\int d^3 x \sqrt{\det g_n}\rho(m^2_{\phi}-m^2_{\psi})+\frac{1}{2} [Tr \log(-\square+\frac{a_n}{r^2})-Tr\log(\partial \!\!\!/+\frac{b_n}{r})]
\end{equation}
If we evaluate $\partial_n S_{\textrm{eff}}(n)\vert_{n=1}$ we would find all the other terms vanishing, since they are  $\mathcal{O}(n-1)^2$ or higher, while the last term vanishes for the trace over $\gamma$ matrices to leading order in $(n-1)$. The remaining terms are
\begin{equation}
\frac{1}{2}[Tr[G_n(a_n)\frac{\partial_n a_n}{r^2}]\vert_{n=1}-Tr\log(-\square)].
\end{equation}
Naively thinking it will be the same order as $a_n$ in $n-1$, to be $O({n-1})^2$, thus giving no contribution to the entropy. However, as we previously showed, there is a non-trivial pole existing in the infrared limit which makes this term the only one contributing to the entropy. And it indeed gives the area law as we will show in the following,
\begin{equation}
\begin{split}
&\partial_n a_n\frac{1}{2}Tr[G_n(a_n)\frac{1}{r^2}]\\
&=\frac{1}{2}\partial_n a_n\int \sqrt{\det g_n} \frac{d^3 x}{r^2}\int \frac{dk_{\perp}}{2 \pi} e^{i k_{\perp}(x_{\perp}-x^{\prime}_{\perp})}\sum_{l}\frac{e^{i l(\theta-\theta^{\prime})}}{2 \pi n}\int^{\infty}_{0}k dk \frac{J_{\nu_l}(k r)J_{\nu_l}(k r^{\prime})}{k^2 +k^{2}_{\perp}}
\end{split}
\end{equation}
Performing the $k_\perp$ integral we get
\begin{equation}
\begin{split}
&\frac{1}{2}\partial_n a_nTr[G_n(a_n)\frac{1}{r^2}]\\
&=\frac{1}{4}\partial_n a_n\int \sqrt{\det g_n}\frac {d^3 x}{r^2}\sum_{l}\frac{e^{il(\theta-\theta^{\prime})}}{2 \pi n}\int^{\infty}_{0} e^{-k \epsilon}dk {J_{\nu_l}(k r)J_{\nu_l}(k r)}\\
&=\frac{1}{4}\partial_n a_n\sum_{l}\frac{1}{2 \pi n}\int^{2\pi}_{0} n d\theta \int^{\infty}_{0}\frac{d r}{r}J_{\nu_l}(k r)J_{\nu_l}(k r)\int^{\infty}_{\infty}dx_{\perp}\int^{\infty}_{0}e^{-k \epsilon}dk \\
&=\frac{1}{4}\partial_{n} a_n\sum_{l}\frac{1}{2 \pi}\int^{2\pi}_{0} d \theta \frac{1}{2\nu_l}\int^{\infty}_{-\infty}dx_{\perp} \frac{1}{\epsilon}
\end{split}
\end{equation}
We find that in the limit $n\rightarrow 1$ the $|l|> 1$ terms vanish since $\partial_n a_n$ is of order $O(n-1)$. However the $l=0$ term remains. Therefore, the leading term in the entanglement entropy is given by
\begin{equation}
S_{EE} =\frac{\hat\alpha L}{2 \epsilon} + \cdots,
\end{equation}
here $\cdots$ could include subleading term in the large area expansion. Also, $\epsilon=\Delta x_\perp$ is the short range cut-off in the $x_{\perp}$ direction and $L$ is the scale of the box in the dimension $x_\perp$, while $\hat \alpha =\lim_{n\to 1} \alpha_n/(n-1)$ .
If we take the solution of the gap equation in (\ref{sloution1}), we find that the area term in the entanglement entropy to leading order in the large $N$ limit admits no dependence on the interaction coupling $g$.

\section{Conclusion} \label{sec:conclude}
Motivated by a lack of computable examples of entanglement entropy of interacting field theories,
we study the $\mathcal{N}=1$ supersymmetric $O(N)$ vector model in $d=3$ near the second order phase transition line, and computed the entanglement entropy of the half space, extracting the leading area term. By considering the entanglement of the half space, whose volume and also boundary area diverge, we have a priori make it almost impossible to extract the subleading universal constant term in the entanglement. Yet, since the area term itself encapsulates in reality most of the quantum entanglement of the ground state, and the physical significance of the emergence of an area term in a local field theory in the first place, we would like to understand whether the variation of the interaction coupling makes any qualitative difference to this leading term. 

It turns out the supersymmetric theory has lots of similarities with the scalar $O(N)$ model at the critical point. The correction of the massless Fermionic propagator in conical space computed perturbatively in $n-1$ has some new divergences whose counter terms we have not been able to pin down uniquely. Nevertheless, this term remains finite at any finite $n-1$, and acquires a $1/(n-1)$ pole enhancement. We solved the gap equations of the system perturbatively in $n-1$, and found surprisingly that the Bosonic mass acquires exactly the same value as in the critical scalar theory found in \cite{Metlitski:2009iyg}, independently of the coupling constant $g$ that can be varied freely along the critical line. An interesting note here is that we found two distinct ways of computing this quantity, by changing the order of limits-- one in which $n-1$ is taken as the smallest scale and expanded first, such that a logarithmic divergence near the conical singularity would arise and call for a localized counter term; and one in which the $r$ integral is done first before the $n\to 1$ limit is taken, as in \cite{Metlitski:2009iyg}. It turns out that the two matches, if the couplings of the counter term is chosen to take its fixed point value, suggesting that the value is robust and unique. Nonetheless, combining with the Fermionic results, we arrive at the leading area term of the entanglement entropy that is only sensitive to the Bosonic mass, and thus independent of the coupling $g$. We suspect this is a large N artifact, and that a 1/$N$ correction should reveal more intricate dependence of the coupling. 

We made other interesting observations along the way. Particularly, we notice the connection between, on the one hand, the equation of the Fermionic green's function in conical space in the presence of a mass term $b_n/r$, and the Dirac equation describing electron in a relativistic hydrogen atom on the other.  The bound states of the relativistic hydrogen atom has been carefully studied and it is a subject discussed in textbooks. A good review can be found for example in \cite{AQMbook}.  These bound state solutions diverges when substituting in the parameters relevant in our problem. However we believe that scattering states should have a sensible interpretation. The connection is to be studied in more detail in future work. 

As mentioned above, the computation of the correction to the Fermionic propagator also allowed the choice of two order of limits, much like the Bosonic ones. In this case, however, the logarithmic divergence is not obviously associated to a local counter term. Our computation has assumed that the final result should agree with that following from the other order of limit. But if in the unpalatable scenario that counter terms with differing subtraction scheme could alter the result, we studied some typical possibilites, and found that unsurprisingly the coupling constant could in fact enter into the area term if such counter term ambiguities do exist. The physics question we asked is unambiguous, and we do not believe such ambiguity could exist, as we demonstrated in the Bosonic case.  Nonetheless, this is an important question -- whether the replica trick does recover uniquely the entanglement entropy of a given wavefunction. 

The subleading universal term also holds key information, along with $1/N$ corrections. We would like to leave these important questions for future investigations.

\appendix

\section{Counter term calculation of Bosonic propagator} \label{D}
In this parrt we show explicitly the calculation of the counter term of boson mass in Eq.(\ref{counterterm}).
\subsection{$-G({\boldsymbol{ r};\boldsymbol{r_1}})c \frac{\delta(r_1)}{r_2}G(\boldsymbol{r_1};\boldsymbol{r})$}
The first order correction in c could be written as 
\begin{equation}
\begin{split}
&-c\int^{\infty}_0\delta(r_1)dr_1\int^{2\pi n}_0d\theta_1\int^{\infty}_{-\infty}dx_1\frac{1}{(4\pi n)^2}\\&(r_0^2+r_1^2-2r_0r_1\cos(\theta_0-\theta_1)+(x_0-x_1)^2)^{-1/2}(r_1^2+r_2^2-2r_1r_2\cos(\theta_1-\theta_2)(x_1-x_2)^2)^{-1/2}\\
=-&\frac{c}{16\pi n}\int^{\infty}_{-\infty}dx_1(r_0^2+(x_0-x_1)^2)^{-1/2}(r_2^2 +(x_1-x_2)^2)^{-1/2}\\
=-&\frac{c}{16\pi n}\int^{\infty}_{-\infty}dx_1(r_0^2+(x_0-x_1)^2)^{-1}(\textrm{take} r_0=r_2, x_0=x_2)\\
=-&\frac{c}{16 n r_0}
\end{split}
\end{equation}

\subsection{$G(\boldsymbol{ r};\boldsymbol{r_1}) c \frac{\delta(r_1)}{r_1}G(\boldsymbol{ r_1};\boldsymbol{r_2}) c \frac{\delta(r_2)}{r_2}G(\boldsymbol{ r_2};\boldsymbol{r})$}
the secound order correction in c takes the form
\begin{equation}
\begin{split}
&\frac{c^2}{(4\pi n)^3}\int^{\infty}_0\delta(r_1)dr_1\int^{2\pi n}_0d\theta_1\int^{\infty}_{-\infty}dx_1\int^{\infty}_0\delta(r_2)dr_2\int^{2\pi n}_0d\theta_2\int^{\infty}_{-\infty}dx_2\\
&(r_0^2+r_1^2-2r_0r_1\cos(\theta_0-\theta_1)+(x_0-x_1)^2)^{-1/2}(r_1^2+r_2^2-2r_1r_2\cos(\theta_1-\theta_2)+(x_1-x_2)^2)^{-1/2}\\
&(r_2^2+r_3^2-2r_2r_3\cos(\theta_2-\theta_3)+(x_2-x_3)^2)^{-1/2}\\
=&\frac{c^2}{(2\pi n)^3}\int^{\infty}_0\delta(r_1)dr_1\int^{2\pi n}_0d\theta_1\int^{\infty}_{-\infty}dx_1\int^{\infty}_0\delta(r_2)dr_2\int^{2\pi n}_0d\theta_2\int^{\infty}_{-\infty}dx_2\int^{\infty}_{-\infty}\frac{dk_1}{2\pi}\int^{\infty}_{-\infty}\frac{dk_s}{2\pi}\int^{\infty}_{-\infty}\frac{dk_3}{2\pi}\\
&e^{i k_1(x_0-x_1)}e^{i k_2(x_1-x_2)}e^{i k_3(x_2-x_3)}K_0[k_1 (r_0^2+r_1^2-2r_0r_1\cos(\theta_0-\theta_1))^{1/2}]\\
&K_0[k_2 (r_1^2+r_2^2-2r_1r_2\cos(\theta_1-\theta_2))^{1/2}]K_0[k_3 (r_2^2+r_3^2-2r_2r_3\cos(\theta_2-\theta_3))^{1/2}]\\
=&\frac{c^2}{(2\pi n)^3}\int^{\infty}_0\delta(r_1)dr_1\int^{2\pi n}_0d\theta_1\int^{\infty}_0\delta(r_2)dr_2\int^{2\pi n}_0d\theta_2\int^{\infty}_{-\infty}\frac{dk_1}{2\pi}e^{i k_1(x_1-x_3)}K_0[k_1 (r_0^2+r_1^2-2r_0r_1\cos(\theta_0-\theta_1))^{1/2}]\\
&K_0[k_1 (r_1^2+r_2^2-2r_1r_2\cos(\theta_1-\theta_2))^{1/2}]K_0[k_1 (r_2^2+r_3^2-2r_2r_3\cos(\theta_2-\theta_3))^{1/2}]\\
=&\frac{c^2}{8\pi n}\int^{\infty}_{-\infty}\frac{dk_1}{2\pi}e^{i k_1(x_1-x_3)}K_0[k_1 r_0]K_0[k_1 0]K_0[k_1 r_3](\textrm{take}\quad r_0=r_3, x_0=x_3)\\
=&\frac{c^2}{32 n r_0}K_0[k_1 0]\\
=&\frac{c^2}{32 n r_0}(-\log k \epsilon+\log 2-\gamma)
\end{split}
\end{equation}
Since $K_0[k 0]$ is divergent,we hereby introduce $\epsilon$ as a point spliting cut-off. The last line is valid as long as $k \epsilon$ is small. The divergence arises as $r_1\rightarrow0, r_2\rightarrow 0$.

\section{Fermionic Green's function and the hydrogen atom}\label{B}
Besides the perturbative calculation for the Fermionic Green's function in the main text, we can actually use the method of mode expansion to express the Green's function as a sum of eigenfunctions, and this method might have possible further use since it will allow us to calculate the Renyi entropy. And we will show that the eigenfunctions have close connection with 3+1d hydrogen atom.

The eigenfunction equation for eigenvalue E is

\begin{equation}\left(\partial \!\!\!/ +\frac{b_n}{r}\right)\psi =E \psi\end{equation}

after changing coordinate into polar coordinate, we get

\begin{equation}
\left(\partial \!\!\!/ +\frac{b_n}{r}\right)=
\left(
\begin{array}{cc}
{\partial_x}+\frac{b_n}{r}& -e^{-\frac{i \theta }{n}}\left(n\frac{\partial _{\theta }}{r}+i \partial _r\right)\\
-e^{\frac{i \theta }{n}}\left(n\frac{\partial _{\theta }}{r}-i \partial _r\right) & -{\partial_x}+\frac{b_n}{r}
\end{array}\right)
\end{equation}

Then we expand using the eigenfunctions for $\theta$ and x, we have

\begin{equation}
\psi=\int \frac{dk_x}{2 \pi }e^{ik_x}\sum _l 
e^{i \left(l+\frac{1}{2}\right) \frac{\theta}{n}}
\left(
\begin{array}{cc}
 e^{-\frac{i \theta}{2}} & 0 \\
 0 & e^{\frac{i \theta}{2}} \\
\end{array}
\right)
\left(
\begin{array}{c}
 \psi _1 (r) \\
 \psi _2 (r)\\
\end{array}
\right) 
\end{equation}

then we get the eigenfunctions for r

\begin{equation}
\left(i k_x+\frac{b_n}{r}-E\right)\psi _1+\left (-i \frac { (2 l +1+n)} {2 n r} - i\partial _r \right)\psi _2=0
\end{equation}

\begin{equation}
\left (-i \frac { (2 l +1-n)} {2 n r} + i\partial _r \right)\psi _1+ \left(i k_x+\frac{b_n}{r}-E\right)\psi _2=0
\end{equation}

Now we introduce dimensionless functions $F(r)=\sqrt{r} \psi_1 $ $G(r)=\sqrt{r} \psi_2 $, the equations become

\begin{equation}
\left(k_x-i\frac{b_n}{r}+i E\right)F-\partial _r G- \frac { 2 l +1} {2 n r} G=0
\end{equation}

\begin{equation}
\left(-k_x-i\frac{b_n}{r}+i E\right)G+\partial _r F- \frac { 2 l +1} {2 n r} F=0
\end{equation}

Comparing with the equations of 3+1d hydrogen atom, whose eigenfunctions are

\begin{equation}
\left(m_c-\frac{\alpha}{r}-\epsilon m_c\right)F-\partial _r G- k G=0
\end{equation}

\begin{equation}
\left(-m_c-\frac{\alpha}{r}-\epsilon m_c\right)G+\partial _r F- k F=0
\end{equation}

Using the well-known result for bound states of hydrogen atom, we have the eigenvalues for the bound states to be

\begin{equation}
E=\frac{i k_x}{\Big{(}1-{{b_n}^2}{\big{(}n-|\frac { 2 l +1} {2 n r}|+\sqrt{|\frac { 2 l +1} {2 n r}|^2+{b_n}^2}\big{)}^{-2}}\Big{)}^{1/2}}
\end{equation}

However, this just means that there are no bound states for the current problem, since now $|E|>k_x$, which is inconsistent with the assumption used to solve for the bound state problem. This example gives the reason why in these kinds of problems we never use bound states to construct the set of complete bases.

On the other hand, calculation and summation for the scattering problem of hydrogen atom has not been fully considered in the literature up to the limited knowledge of the writers. So we end our discussion here before we can go further for the problem of getting the whole spectrum and Renyi entropy. So we went step back to take the second best and calculate the n-1 purturbation to get the entanglement entropy 
in our main text.

\section{Correction of the Fermionic Green's function} \label{B}

In this section we may show that the logarithmic divergence in Eq.(\ref{fcorrectione}) can be subtracted in the form of a pole in $1/(n-1)$.

We start with
\begin{equation}
\begin{split}
&{\chi _n}{\rm{(}}{{\rm{b}}_n}{\rm{) = }}tr[{\partial \!\!\!/^{ - 1}} - {\partial \!\!\!/^{ - 1}}(\delta ){\partial \!\!\!/^{ - 1}}] = tr[{\partial \!\!\!/_{\rm{r}}}G(r,r'')(\frac{{{b_n}}}{{r''}}){\partial \!\!\!/^{ - 1}}(r'',r')] = tr[G(r,r'')\overleftarrow {{{\partial \!\!\!/}_{\rm{r}}}} (\frac{{{b_n}}}{{r''}}){\partial \!\!\!/^{ - 1}}(r'',r')]\\
 =& tr[ - G(r,r'')\overleftarrow {{{\partial \!\!\!/}_{{\rm{r''}}}}} (\frac{{{b_n}}}{{r''}}){\partial \!\!\!/^{ - 1}}(r'',r')] = tr[G(r,r'')\left( {(\frac{{{b_n}}}{{r''}}){{\partial \!\!\!/}^{ - 1}}(r'',r')} \right)] = tr[G(\partial \!\!\!/\delta \partial \!\!\!/){\partial \!\!\!/^{ - 1}} + G(\delta \partial \!\!\!/)\partial \!\!\!/{\partial \!\!\!/^{ - 1}}]\\
 =& 2\frac{{{b_n}}}{r}({G_n}({m_\psi } = 0,\gamma  = 1/2)) - {b_n}\int {\frac{{dxd\theta dr}}{{{r^2}}}} G_n^2({b_n},\gamma  = 1/2) + {b_n}\int d \theta dx\left[ {\frac{{G{{({m_\psi } = 0,\gamma  = 1/2)}^2}}}{r}} \right]\Big{|}_{r \to 0}^{r \to \infty }
\end{split}
\end{equation}
where the last surface term can be shown to be zero.

Now we focus on the second term above 

\begin{equation}
\begin{split}
&\int{\frac{{{G^2}}}{{{r^{\prime 2}}}}} {d^3}{x^\prime} = \int_0^{2n\pi } d {\theta ^\prime }\int d {x^\prime }\int_0^\infty  {\frac{{d{r^\prime }}}{{{r^{\prime 2}}}}} \sum\limits_{l =  - \infty }^{l = \infty } {\frac{{{e^{i{\nu _l}(\theta  - {\theta ^\prime })}}}}{{2\pi n}}} \sum\limits_{{l^\prime } =  - \infty }^{{l^\prime } = \infty } {\frac{{{e^{i{\nu _{{l^\prime }}}({\theta ^\prime } - \theta \prime \prime )}}}}{{2\pi n}}} \\
&\int {\frac{{d{k_ \bot }}}{{2\pi }}} {e^{i{k_ \bot }(x - {x^\prime })}}\int {\frac{{d{k^\prime}_ \bot }}{{2\pi }}} {e^{i{k^\prime}_ \bot ({x^\prime } - {x^{\prime\prime}})}}\int_0^\infty  {\frac{{{J_{{\nu _l}}}(kr){J_{{\nu _l}}}(k{r^\prime })}}{{{k^2} + k_ \bot ^2}}} kdk\int_0^\infty  {\frac{{{J_{{\nu _l}}}({k^{\prime}}{r^{\prime\prime}}){J_{{\nu _l}}}({k^{\prime}}{r^\prime })}}{{{k^{\prime}}^2 + {k^{\prime}}_ \bot ^2}}} {k^\prime}d{k^\prime}\\
& = \sum\limits_{l =  - \infty }^{l = \infty } {\frac{1}{{4\pi n}}} \int_0^\infty  {\frac{{d{r^\prime }}}{{{r^{\prime 2}}}}} \int_0^\infty  d k\int_0^\infty  d {k^\prime }\frac{{{J_{{\nu _l}}}(kr){J_{{\nu _l}}}({k^\prime }{r^{\prime \prime }}){J_{{\nu _l}}}(k{r^\prime }){J_{{\nu _l}}}({k^\prime }{r^\prime })}}{{k + {k^\prime }}}\\
& = \sum\limits_{l =  - \infty }^{l = \infty } {\frac{1}{{4{\pi ^{3/2}}n}}} \int_0^\infty  d k\int_0^1 {\frac{{{t^{{\nu _l}}}}}{{1 + t}}} \frac{{\Gamma ( - 1/2 + {\nu _l})}}{{\Gamma (1 + {\nu _l})}}{}_2{F_1}( - 1/2, - 1/2 + {\nu _l};1 + {\nu _l},{t^2}){J_{{\nu _l}}}(kr){J_{{\nu _l}}}(ktr)\\
& = \sum_{l=-\infty }^{l =\infty} {\frac{1}{{8{\pi ^2}n{r^2}}}} \frac{1}{{\nu_l^2 - 1/4}} = \frac{{\tan (\frac{\pi }{2}n)}}{{4\pi {r^2}}} 
\end{split}
\end{equation}

When we investigate the limit $n\rightarrow 1$, we expand the expression to get that
\begin{equation}
\frac{{b_n\tan (\frac{\pi }{2}n)}}{{4\pi {r^2}}} =-\frac{b_n}{2\pi^2r^2}\frac{1}{(n-1)}+\frac{b_n(n-1)}{24 r^2}+O(n-1)^2
\end{equation}
We find that the result is in correspondence to Eq.(\ref{fcorrectione}) in that the coefficient of the divergence as well as the O(1) term is exactly the same.

\section{Other solutions of the gap equation?}

In the case of bosons, we were able to pin down a fixed point value of these couplings of counter term, leading to an answer that appears to be robust against the choice of different normalization schemes. For fermions, this issue is not well understood, and so here we explore the consequence should any scheme dependence in the gap equation actually survive. Suppose we subtract the leading term $\frac{b_n}{2\pi^2 r^2}$ in the gap equation by hand without referring to any fixed point value of a counter term. In that case, they become a set of homogeneous equations, with both $\alpha_n$ and $b_n$ poportional to $(n-1)$.  So generically we write
\begin{equation}\label{alphabet}
\alpha_n=(n-1)\alpha(g_{\phi_2},g_\psi),\quad b_n=(n-1)b(g_{\phi_2},g_\psi)
\end{equation}

in which a and b are functions that depend only on the coupling constants.

Two sets of the solutions to Eqs.(\ref{alphabet}) are
\be
\begin{array}{cc}
\alpha =\frac{9 g_{\phi_2}^2+2 g_{\psi}+2 \sqrt{432 g_{\phi_2}^2+g_{\psi} \left(g_{\psi}+192\right)}}{768-36 g_{\phi_2}^2},& b=\frac{g_2 \left(g_{\psi}+\sqrt{432 g_{\phi_2}^2+g_{\psi} \left(g_{\psi}+192\right)}+96\right)}{48 \left(3 g_{\phi_2}^2-64\right)}\\
\alpha=\frac{9 g_{\phi_2}^2+2 g_{\psi}-2 \sqrt{432 g_{\phi_2}^2+g_{\psi} \left(g_2+192\right)}}{768-36 g_{\phi_2}^2},& b= -\frac{3 g_{\psi}}{g_{\psi}+\sqrt{432 g_{\phi_2}^2+g_{\psi} \left(g_{\psi}+192\right)}+96}
\end{array}
\ee

A noteworthy feature of the solution is that the solution would be divergent at given value $g_{\phi_2}=8/\sqrt{3}$, regardless of other parameters. This indicates a possible phase transition at this specific value. However, another set of solution is regular whatever values of the coefficents take. Whether such a scenario would ever arise will be explored in future work.

\section*{Acknowledgements} 
We thank A Bhattacharyya , C-T Ma and M. Metlitski for useful discussions. We thank particularly S. Sachdev and W. Witczak-Krempa for sharing with us techniques to use Pauli-Villars regularization in conical space, and for sharing parts of their notes on their forth-coming paper. 
We also thank M. Smolkin for a careful read of our manuscript and for his many valuable comments. LYH would like to acknowledge support by the Thousand Young Talents Program, and Fudan University. 

\emph{hi}

\bibliographystyle{utphys}

\end{document}